\documentclass[12pt]{iopart}
\usepackage{iopams}  
\expandafter\let\csname equation*\endcsname\relax
\expandafter\let\csname endequation*\endcsname\relax
\usepackage{amsmath,amssymb,amsfonts} 
\usepackage{graphicx} 
\usepackage{subfigure}

\usepackage{commands}  % Cordes abbreviations and macros *** Need "commands.sty"

\newcommand{\be}{\begin{equation}} 
\newcommand{\ee}{\end{equation}}

\newcommand{\bb}{\begin{bmatrix}}
\newcommand{\eb}{\end{bmatrix}} 
\newcommand{\Deff}{D_{\rm eff}}
\newcommand{\gae}{\lower 2pt \hbox{$\, \buildrel {\scriptstyle >}\over {\scriptstyle
\sim}\,$}}
\newcommand{\lae}{\lower 2pt \hbox{$\, \buildrel {\scriptstyle <}\over {\scriptstyle
\sim}\,$}}
\newcommand{\us}{$\mu$s}
\newcommand{\ev}[1]{E\left\{#1\right\}}

\begin{document}

\title[Effects of the ISM on Detection of Low-frequency Gravitational Waves]{Effects of the Interstellar Medium on Detection of Low-frequency Gravitational WavesÊ}

\author{Dan~Stinebring}

\address{Oberlin College, Dept.~of Physics and Astronomy, Oberlin, OH 44074}

\ead{dan.stinebring@oberlin.edu}

\begin{abstract} 
Time variable delays due to radio wave propagation in the ionized interstellar medium are a substantial source of error in pulsar timing array efforts.
We describe the physical origin of these effects, discussing dispersive and scattering effects separately. Where possible, we give estimates of the magnitude of timing errors produced by these effects and their scaling with radio frequency.
Although there is general understanding of the interstellar medium propagation errors to be expected with pulsar timing array observations, detailed comparison between theory and practice is still in its infancy, particularly with regard to scattering effects.
\end{abstract}

%XXXXX
%\input{section_introduction}
\section{Introduction}
As is clear from other articles in this volume, the success of the pulsar timing array (PTA) effort depends on
measuring pulse arrival times to an accuracy of tens of nanoseconds over five years or more.
Since the radio waves from pulsars travel through hundreds or thousands of light years of interstellar space before arriving at our telescopes,
we need to accurately correct for timing delays caused by this passage.
It is the ionized component of the interstellar medium (ISM) that interacts strongly
with radio waves and causes a myriad of effects that are primarily deleterious to pulsar
timing.

\begin{figure}[h!]
\begin{center}
     \includegraphics[scale=0.2, angle=0]{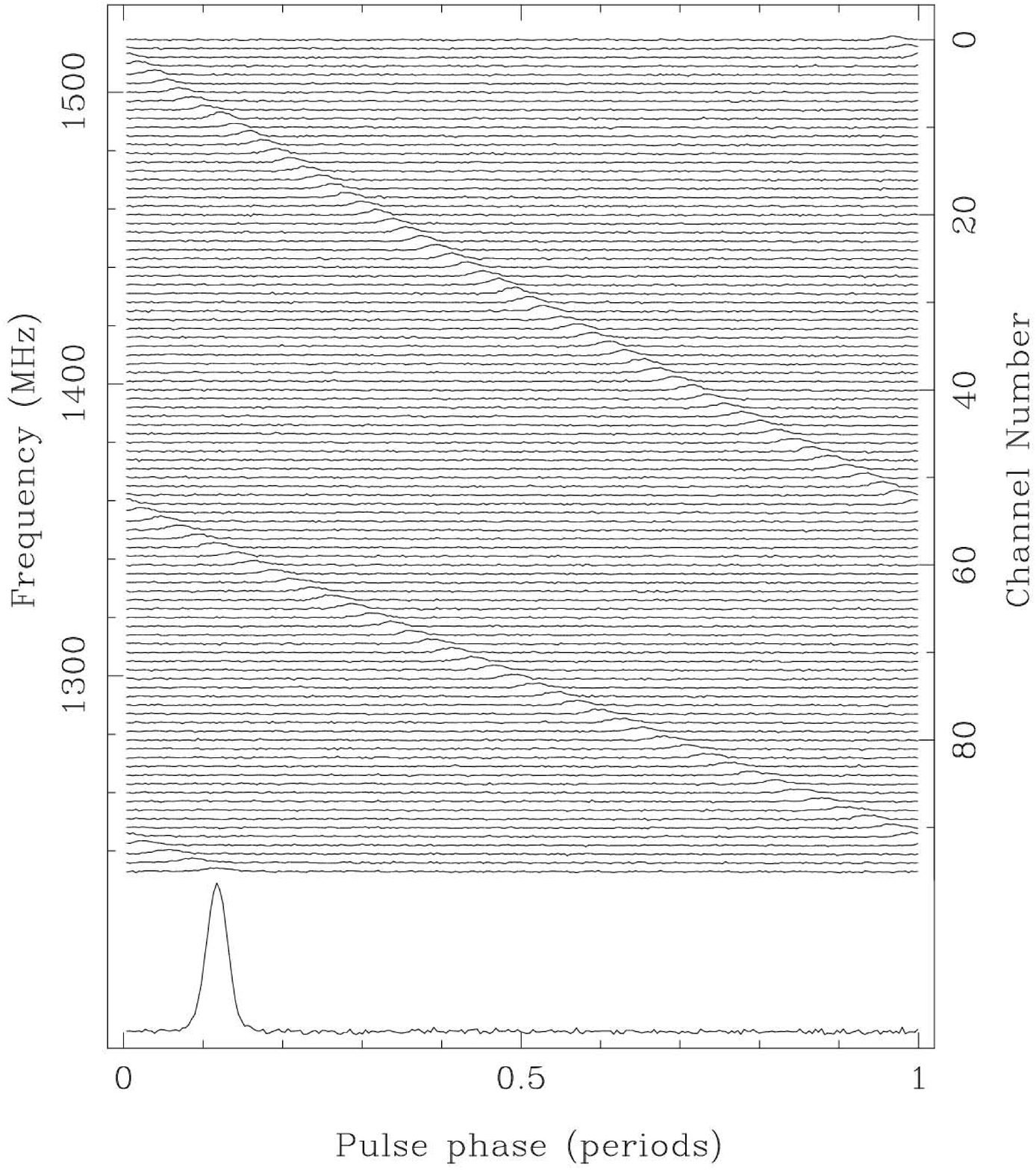}
   \caption{Dispersive delay for a pulsar observation centered at 1380 MHz.  Over this fairly wide fractional bandwidth the dispersive slope changes significantly. Pulses are folded modulo the pulsar period causing the apparent discontinuity around channel 52.From \cite{lk05}
    	\label{fig:dispersion}
   }
\end{center}
\end{figure}

The two broad classes of propagation effects are those due to {\em dispersion} (figure~\ref{fig:dispersion}), which would be
present in a homogeneous medium and {\em scattering}, which is due to
spatial inhomogeneities in the medium.
% Lazio - frequency dependent refractive index?
All effects are traceable to the oscillations of the plasma induced by the passage
of the e-m wave of frequency $\nu$.
Dispersion arises when the radio waves, which span a range of
frequencies, interacts with the plasma.
The frequency-dependent group velocity is $v_g = c \sqrt{1 - \nu_p^2/\nu^2}$, where
the plasma frequency is $\nu_p = \sqrt{n_e e^2/(\pi m_e)} \approx 
(10^4~{\rm Hz}) \sqrt{n_e / 1 {\rm cm}^{-3}}$,
with the electron density $n_e \approx 0.03 {\rm cm}^{-3}$ in the Galactic plane.
When the radio wave propagates through a column of free electrons
characterized by the dispersion measure DM$= \int{n_e ds}$, it will be delayed by
a time equal to \cite{lk05} 
\be
t_{\rm DM} (\nu) = \frac{e^2}{2\pi m_e c^3 \, \nu^2}\; {\rm DM}
\approx (4.15\;{\rm ms})\frac{\rm DM}{\nu_{\rm GHz}^{2}},
\label{eqn:dm}
\ee
where, in the last expression, DM is in the standard units of pc~cm$^{-3}$.

% Lazio - added the following paragraph.
A homogeneous ionized ISM would produce dispersion; in fact, an ionized medium that has variations in $n_e$ only along the propagation direction produces dispersive effects that are completely correctable.
However, spatial
inhomogeneities  transverse to the line of sight produce multi-path propagation effects
or scattering. 
There are multiple
effects but the most important for this discussion are intensity scintillations and pulse
broadening.

\begin{figure}[h!]
%\begin{center}
      \hfill \includegraphics[scale=0.45, angle=0]{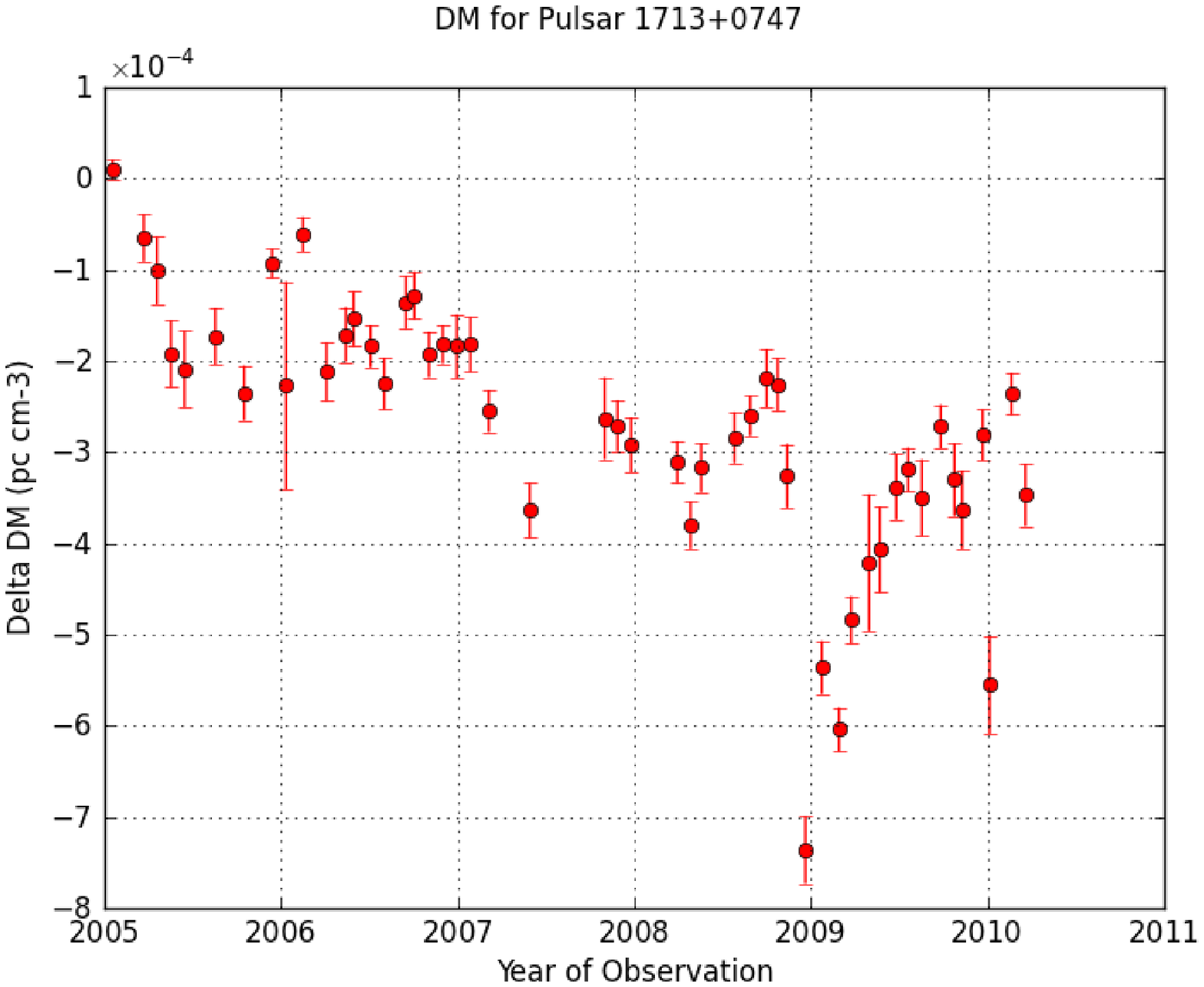}
      \hfill \includegraphics[scale=0.50, angle=0]{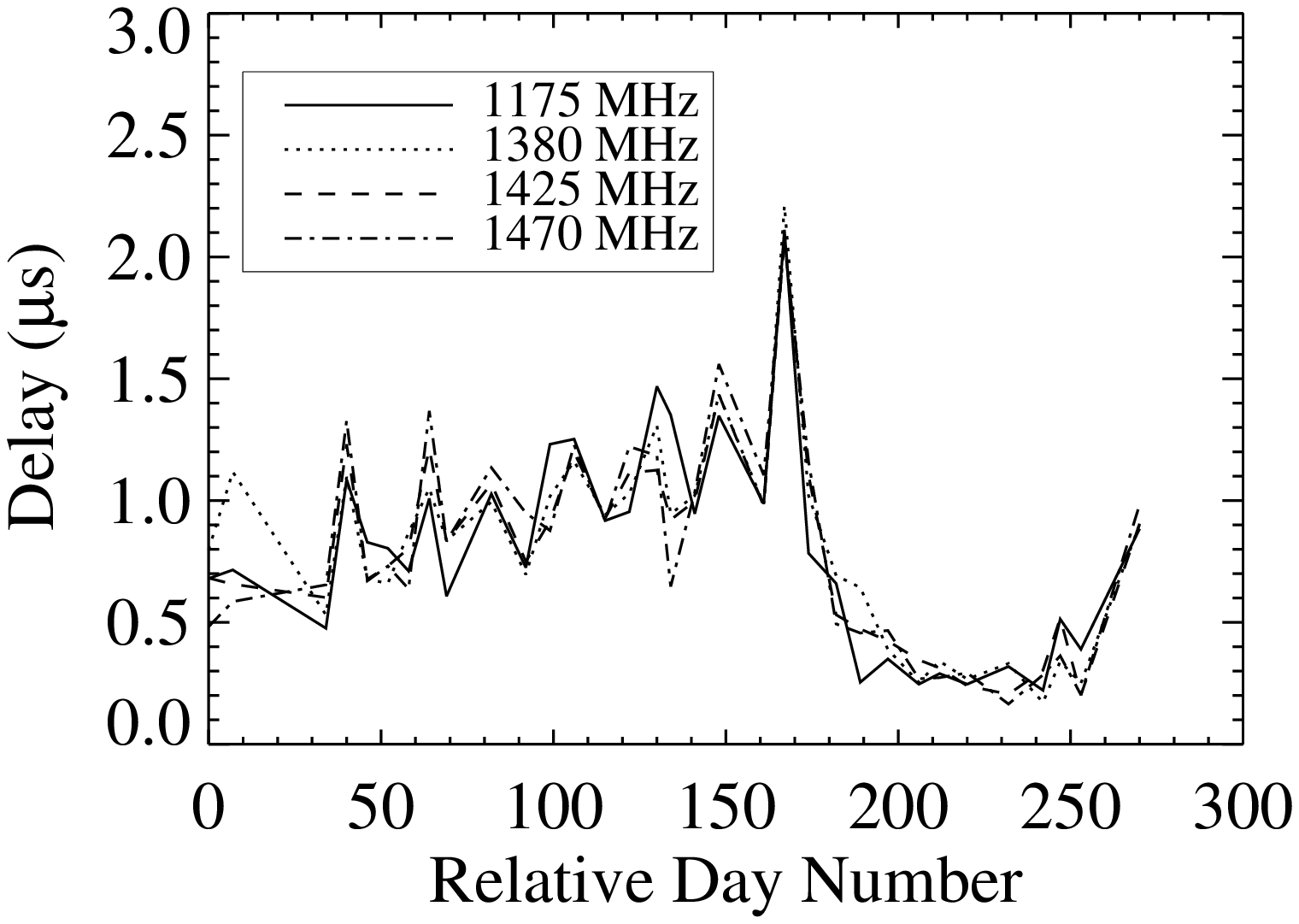}
   \caption{(Left) Dispersion measure variations of PSR~J1713+0747 for 
   approximately five years from dual-frequency NANOGrav observations
   \cite{dfg+13}. A fluctuation of DM by $1.0 \times 10^{-4}$~pc~cm$^{-3}$ produces
   an arrival time variation of 0.41~\us\ at an observing frequency of 1~GHz.
   The abrupt dip and recovery near the start of 2009 was corroborated by
   other PTAs\cite{kcs+13}, which also monitor this pulsar.
   (Right) Scattering delay from PSR~B1713+37 as determined indirectly
   from the power distribution in the secondary spectra\cite{hs08}.
   The four curves have been shifted vertically using 1380~MHz as a reference frequency. 	\label{fig:demHS08f2}
   }
%\end{center}
\end{figure}

If pulsars, the Earth, and the interstellar medium were stationary the constant time delay produced by both dispersion and scattering would present no
problem for PTA efforts. 
However, pulsars are high velocity objects, in general \cite{lk05}, and the Earth and
ISM motions cannot be neglected in all cases.
This motion gives rise to variable delays in the time of arrival (TOA) of pulses which, if uncorrected or improperly corrected, produce systematic errors that can swamp any GW signal.
Figure~\ref{fig:demHS08f2} shows examples of time variable dispersion measure and time variable scattering delay -- including occasional rapid fluctuations (timescales of days) -- that require careful detection and mitigation.
\vspace{0.1in}   % there is a collision with the page heading. Bounding box adjustment?

%XXXXX
%\input{section_ISM}
\section{The ionized interstellar medium}
\label{sec:ism}
\begin{figure}[htb]
\begin{center}
        \includegraphics[scale=0.9, angle=0]{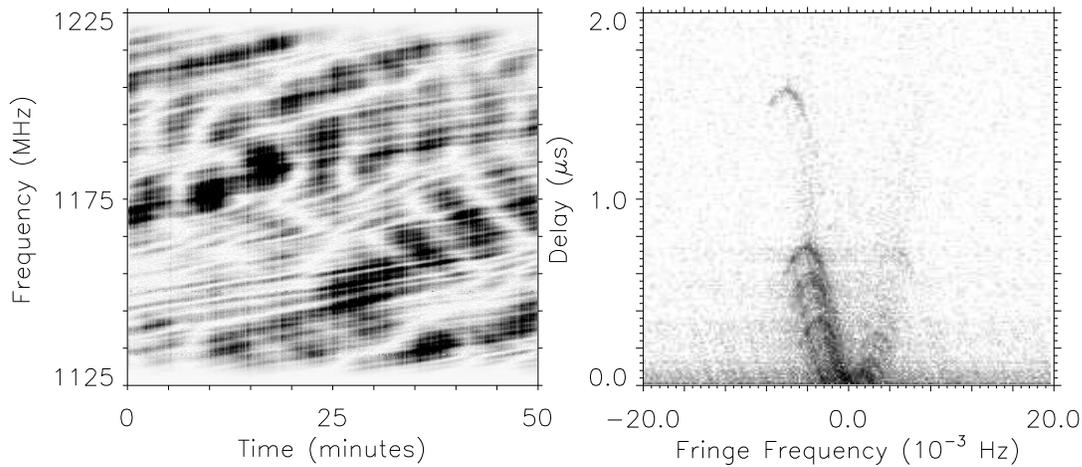} 
   \caption{(Left) A dynamic spectrum of PSR~B0834+06 observed at Arecibo. Flux density
   is indicated using a linear grayscale with dark representing the strongest power. The vertical striping is caused by a small number of pulses per 10~s resolution element. The criss-cross pattern is somewhat unusual and is caused by interference effects when flux arrives from a few discrete directions. (Right) The secondary spectrum (2d power spectrum of the dynamic spectrum) of the same data. The overall parabolic (scintillation) arc indicates that the majority of the scattering occurs in a relatively confined portion along the line of sight. The inverted parabolic ``arclets" are due to discrete scattered rays that interfere strongly with the central part of the pulsar image. Power is represented using a logarithmic grayscale (darkest is the strongest power), and the value along the ordinate indicates the overall delay of the scattered ray relative to the central one.
   From \cite{crsc06}.
    	\label{fig:secspec}
   }
\end{center}
\end{figure}

Before proceeding we set the basic scenario and define some important quantities.
In figure~\ref{fig:secspec} we see the effects of multi-path scattering and the ensuing interference because of the very compact size of pulsars as radio sources.
Both a dynamic spectrum and its 2d power spectrum (the ``secondary spectrum") are shown.
The dynamic spectrum is comprised of {\em scintles}, which are isolated islands of power in the frequency-time plane.
They arise in the time dimension because the random interference pattern at the 
location of the Earth has a characteristic spatial scale and this becomes a 
characteristic time scale due to the relative motion in the problem.
The scintles have a characteristic extent in frequency  because the coherent rays
arriving at the Earth have a range of extra scattering delays of $\sim \tau$ and the
uncertainty principle gives rise to a  diffractive bandwidth $\Delta\nu_d \sim (2\pi\tau)^{-1}$ .
See \cite{cr98} for a careful definition of all quantities and a discussion of how
they depend on the spectrum of density inhomogeneities in the medium and
the spatial distribution of scattering material along the line of sight.
It is often valid to assume a ``frozen flow" approximation since pulsar transverse velocities
are typically $> 100$~km/s, the Earth's contribution is $\sim 30$~km/s, whereas the 
thermal speed of the ISM is $\sim 10$~km/s.

From secondary spectra similar to this
\cite{crsc06,smc+01,shm+03,hsb+03,hsa+05,pmsr05,ps06a,
sti07,hs07,rscg11},
there is evidence that the majority of scattering along many lines of sight to nearby pulsars takes place in discrete screens, by which we mean a relatively small fraction 
(upper limits $\lae$ a few percent) of the line of sight (LOS) distance.
This simplifies a description and analysis of effects, and we use this approximation below, although the apparently common case of multiple screens along the LOS \cite{ps06a} has not been properly analyzed.
Following \cite{cs10} and many other studies we indicate the main scattering angle in the screen as $\theta_d$ since it results from diffraction from small scale ($10^5 {\rm m}< l_d < 10^9{\rm m}$) irregularities.
All quantities are approximate; see \cite{ric90} for exact definitions. 
At an effective distance $D_{\rm eff}$ from the scattering screen \cite{cr98,cs10} this represents a patch size often referred to as the refractive scale
$l_r \approx \theta_d D_{\rm eff}$ since irregularities in the screen of this size will act to
refract the entire ray bundle \cite{rcb84,ssh+00}.
As viewed from Earth, the amount by which the centroid of the ray bundle is refracted is denoted as $\theta_r$.
There is no guarantee that the ray bundle picture is appropriate for all cases since it is known from spectral studies that pulsars occasionally undergo episodes of {\em multi-imaging} in which the image, if it could be resolved, breaks into multiple pieces 
\cite{crsc06,cs10,ebf+70,ra82,hwg85,cw86,wc87,rlg97,lr99}.

For further study of scintillation in the ionized interstellar medium see \cite{ric90} for a more formal and definitive treatment; \cite{nar92a} for another excellent, physics-oriented approach; good introductory discussions in \cite{lk05,lgs12}; and an overview of interstellar turbulence in \cite{es04,se04}.

%XXXXX
%\input{section_dm}
\section{Correcting for Dispersion Measure Variations}
\label{sec:dm}
%You et al 2007; Keith et al 2013.
It is clear from (\ref{eqn:dm}) that uncorrected or improperly corrected DM
variations will seriously bias GW detection efforts.
In an important recent study, Keith et al.\ \cite{kcs+13}, building on previous
work \cite{yhc+07}, show that some algorithms for
dispersion measure correction also remove gravitational wave
power from the signal.
As part of the PPTA effort they employ a technique that includes
a common mode term in addition to the standard $\nu^{-2}$ term \cite{kcs+13}:
\be
t_{{\rm OBS}} = t_{\rm CM} + t_{\rm DM} (\lambda / \lambda_{\rm REF} )^2, 
\ee
where $t_{\rm CM}$ is the wavelength-independent (common mode) delay, 
$t_{\rm DM}$ is the dispersive delay at some reference wavelength $\lambda_{\rm REF}$,
and all quantities represent {\em residuals} after a basic timing solution has been 
subtracted.
They show that not including the common mode term will result in a bias in
the determined dispersion measure.
Much more importantly, that bias arises because a portion of the common
mode  (e.g.\ an achromatic GW signal) will be absorbed into the chromatic
fit for DM.
%As a simple example, consider an observation at two observing
%wavelengths: $\lambda_1 = 1$~m and $\lambda_2 = 2$~m. 
%Introduce a signal with $ t_{\rm CM} = 1$~\us\ and $ t_{\rm DM} = 1$~\us\
%at $\lambda_{\rm REF} = 1$~m so that we have observations 
%$t_{{\rm OBS},1} = 2$~\us\ and $t_{{\rm OBS},2} = 5$~\us.
%Then, a solution with a common mode term would properly recover
%the two parameters $ t_{\rm CM}$ and $ t_{\rm DM}$, but a DM only
%determination would produce a biased estimate of 
%$ t_{\rm DM} = 1.29\ \mu$s and absorb a portion of the common mode signal,
%illustrating the danger of ignoring the common mode term.

\begin{figure}[h!]
\begin{center}
      \includegraphics[scale=0.7, angle=0]{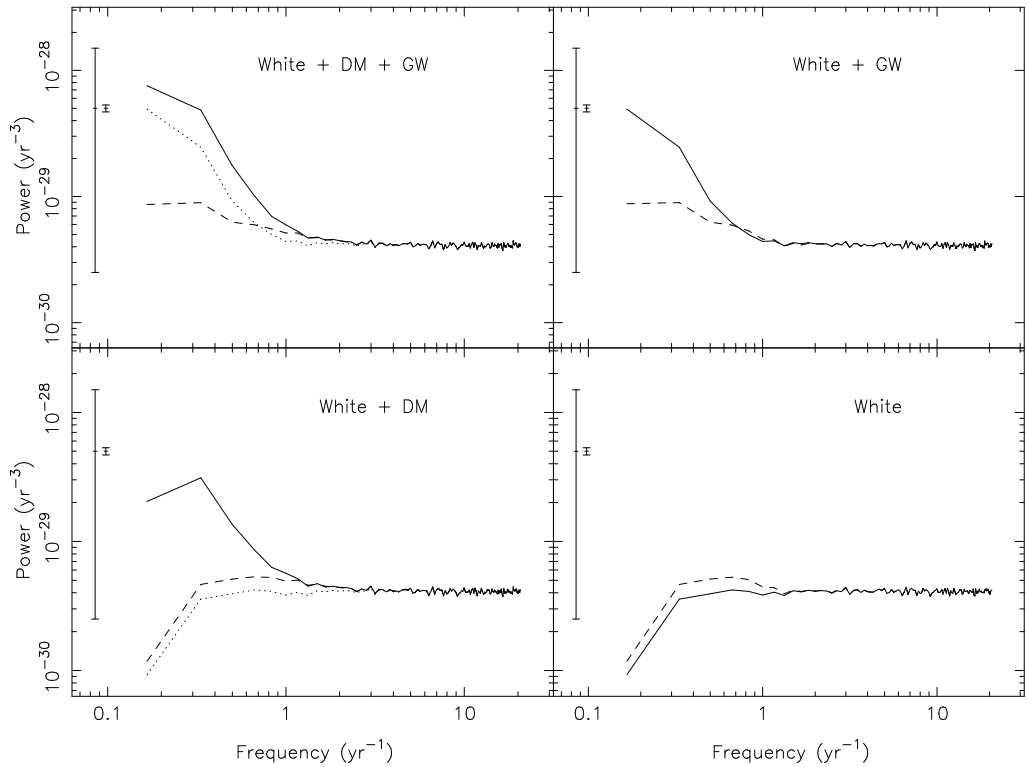}
   \caption{
   Average power spectra of pre- and post-correction timing residuals with four combinations of signals. The solid line shows the pre-correction spectrum and the dashed line shows the post-correction spectrum. For the two cases in which DM fluctuations are introduced, the dotted line indicates the pre-correction power level without DM fluctuations.
The fitting routine uses the DM$(t)$ interpolated fitting routine, without fitting a common-mode signal. See Keith et al.\ \cite{kcs+13} for further details and a comparison with a trial in which a common mode signal is present in the fit.   
	\label{fig:Keith}
   }
\end{center}
\end{figure}

Keith et al. \cite{kcs+13} go on to explore this  using a series of simulations 
that are increasingly realistic.
Results of one simulation are shown in figure~\ref{fig:Keith} in which they do not
include a common mode term in the fit.
In the upper two
panels they include gravitational waves (GW) in
the stochastic signal present in the realizations.
The dashed line, which shows the post-fit power spectrum over
1000 realizations, is strongly suppressed at low frequencies because
achromatic gravitational wave power has been absorbed in the chromatic
fit for $\rm{DM} (t)$.
In a comparison 
(Figure~3 of \cite{kcs+13}) they show that this suppression of the
gravitational wave signal does not occur when a common mode term 
is included in the correction.
An inevitable consequence of including the common mode term
is that a portion of the white noise present in the data ends up
in the post-fit residuals.
This is a consequence of using the least-squares estimator, which is the minimum
variance {\em unbiased} estimator.

% wording from David Nice %%%% June 14, 2013 email message %%%%
The NANOGrav collaboration accounts for the covariance between dispersion measure variation and achromatic red noise at a different stage in their analysis pipeline. 
In  \cite{dfg+13}, they employed a DM correction scheme (called DMX in the TEMPO timing package) that introduces $N_{\rm epoch}$ independent DM values ($DM_i$) for that many (nearly simultaneous) dual-wavelength observations. In the pulse time series analysis, the $DM_i$ values are fit simultaneously with other pulse timing model parameters (rotation rate, spin-down, position, etc.). 
The timing analysis package iteratively calculates barycenter-corrected TOAs, which serve the same function as common mode (``achromatic") terms. 
In  \cite{dfg+13}, the effect of the $DM_i$ fits on the measurement of a gravitational wave limit was accounted for by considering covariance between the basis functions of the $DM_i$ parameters in the timing solution (along with all other timing fit parameters) and putative achromatic gravitational wave signals.
More details of this method of accounting for inclusion of DMX parameters are given in
\cite{vl13,esv13}; also, see the papers by J.~A.~Ellis and X.~Siements et al. in this volume.

The European Pulsar Timing Array (EPTA) is actively developing a DM correction scheme as well. 
Some details were presented at the IPTA meeting in Krabi, Thailand in June~2013, and a full exposition will appear shortly with K.~J.~Lee as lead author.
In broad outline,
they develop a new method inspired by the maximum likelihood estimator.
Compared to other approaches, their  technique utilizes the information that
the dispersion measure variation is temporally correlated among adjacent observations. 
There are two major steps in their method: (i) they measure the power-law spectrum of
dispersion measure using a time-domain spectral analysis, and
then (ii) they construct the linear optimal filter to extract its waveform. 
According to them, this two-step
method  retains good time resolution to short timescale dispersion variation compared 
to a polynomial  fitting algorithm and is able to handle 
(inevitable) irregularly sampled data without  interpolation
because all the signal analyses are performed in the time domain.

\begin{figure}[h!]
\begin{center}
      \includegraphics[scale=0.7, angle=0]{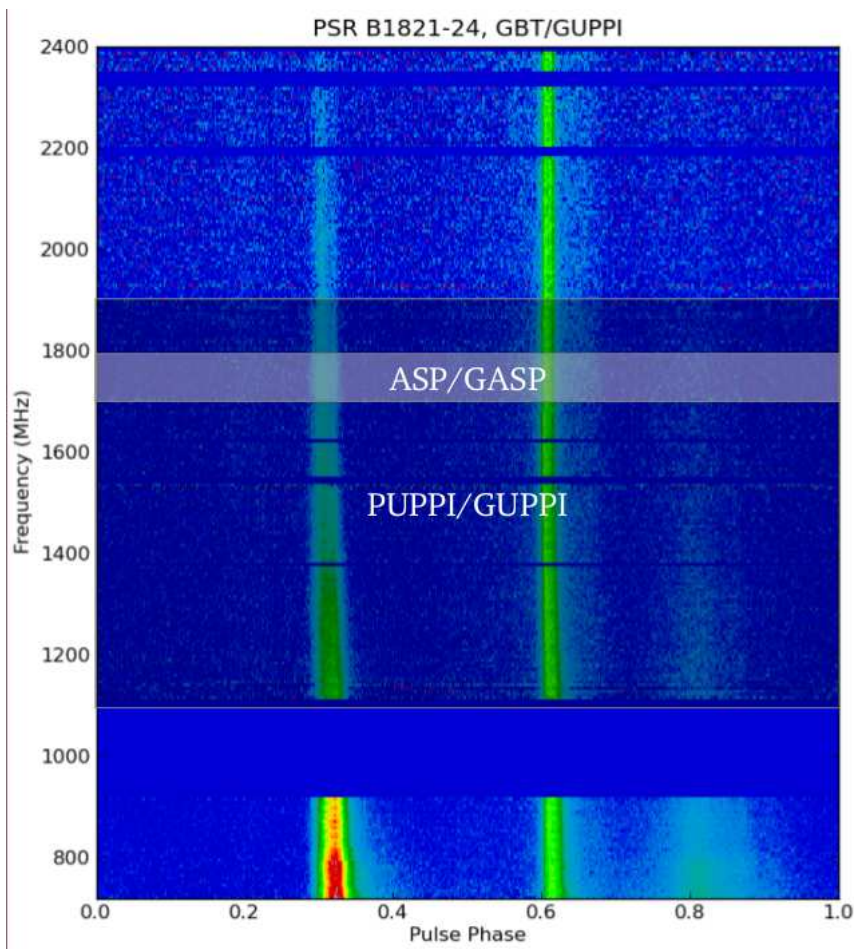}
   \caption{A composite of observations of the globular cluster pulsar
   M28A taken at three bands with the GUPPI wideband spectrometer.
   The bandwidth of the previous generation spectrometers (ASP/GASP, 64~MHz)
   is shown in comparison to the PUPPI/GUPPI instruments (800~MHz).
   (The narrow horizontal stripes are due to radio frequency interference excision, 
   and the gap in
   data between 950 -- 1100~MHz is due to receiver coverage limitations.)
   Around 800~MHz the pulsar profile clearly becomes triple, and the increased
   scattering delay at low frequencies becomes apparent.
   Figure credit: Paul Demorest and Scott Ransom.  
	\label{fig:M28A}
   }
\end{center}
\end{figure}

% discussion of Pennucci wideband timing technique
\vspace{0.2in}
\noindent {\em Wideband Observing Considerations}~~~
As detailed elsewhere in this volume, each PTA has its own particular
resources and limitations. Members of the NANOGrav team have led
the development of two wideband spectrometers, PUPPI and GUPPI,
sited at Arecibo and Green Bank, respectively.
These spectrometers, 
based on CASPER and compute cluster design,
can coherently dedisperse a signal over
800~MHz bandwidth, are extremely versatile,
and have become the production spectrometers for the collaboration.
Their use raises important new questions for how best to 
determine arrival times as is discussed by
Lommen and Demorest in this volume.
The advent of large fractional bandwidths at
frequencies of 1~GHz and above raises new opportunities
and challenges for DM determination (and higher order ISM
effects) as can be seen in
figure~\ref{fig:M28A}, which is an overlay of observations taken
at three different bands with the PUPPI/GUPPI instantaneous
bandwidth highlighted.
Although each $\sim 1.5$~MHz frequency slice has been coherently dedispersed using
the same DM value, the extra time delay due to multi-path scattering becomes evident
at the low frequencies, and even the assumption that one DM value applies over the entire observed range has to be questioned as discussed briefly in section~\ref{sec:scattering}.
These issues and the obvious profile evolution as a function of time are being
explored as central R\&D questions within the various PTAs, which are also moving to wideband observing strategies.

%XXXXX
%\input{section_scattering}
\section{Correcting for Multi-path Scattering Effects}
\label{sec:scattering}
\begin{figure}[h!]
\begin{center}
      \includegraphics[scale=0.60, angle=0]{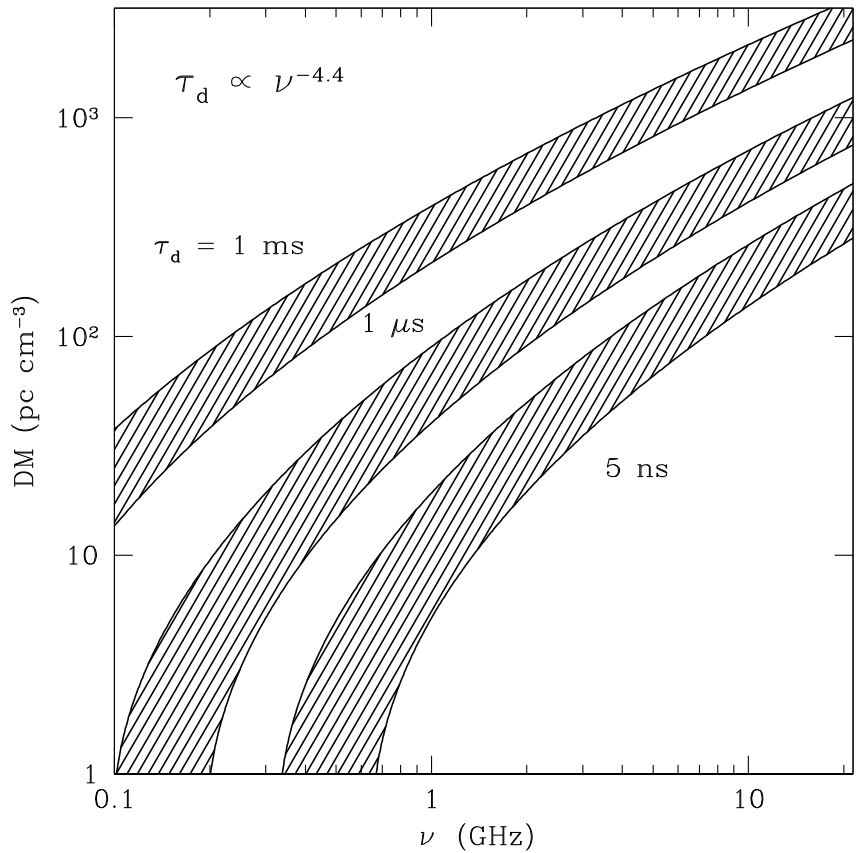}
   \caption{
	Contours of constant pulse broadening time using an empirical
	fit to $\tau_d$ including $\pm 1.0\sigma$ deviations from the fit\cite{cs10}.  
	A $\taud\propto \nu^{-4.4}$
	scaling with frequency is used.  This scaling is steeper
	than appropriate for some high DM objects but is typical for
	low-DM pulsars like the ones that predominate in the PTA source
	lists. From \cite{cs10}.  
	\label{fig:taud}
   }
\end{center}
\end{figure}

% evidence that "thin screen" is appropriate for scattering; thin scintillation arcs
% refer heavily to CS10; use "shift approximation" language
% look at the Cordes comment on minimum delay PDF solutions
% refer to cwd+90 (Cordes 1990; B1937+21 paper)
% Cordes refers to weak, strong, and super-strong scattering
Many of the ISM scattering delays were identified by Foster and Backer
\cite{fb90}. We follow their framework here, updated and with notation
consistent with Cordes and Shanon \cite{cs10}, who identify 6 interstellar scintillation terms:
\be
\dtISS =  
	 \dtPBF 
	+ \dtAOA
	+ \dtAOABary
	+ \dtPBFRISS
	+ \dtPBFDISS
	+ \dtDMnu.
\label{eq:dtiss}
\ee
All of these quantities are time variable, which is the crux of the problem.
The first of these,  $\dtPBF$, are fluctuations associated with the pulse broadening function (PBF), which results from multi-path scattering.
We will loosely refer to the multiple ray paths as a ray bundle, although a full wave propagation formalism \cite{cfrc87,crg+10}
is best employed in simulations of the propagation.
The image on the sky is distorted and wanders, which gives rise to 
time variable angle-of-arrival (geometric) delays $ \dtAOA$.
The changing location of the effective position of the source then
gives rise to an additional error when the TOA is improperly corrected to the solar system
barycenter, $\dtAOABary$.
Diffractive and refractive modulation of the propagation alter the 
PBF and are denoted as $\dtPBFDISS$ and
$\dtPBFRISS$, respectively.
Finally, the ray bundle itself is frequency dependent, and this is
encapsulated in $\dtDMnu$.
\vspace{0.1in}
Using an empirical model of scattering as a function of DM, Cordes and Shannon \cite{cs10} plot scattering time as a function of observing
frequency and DM in Figure~\ref{fig:taud}.

Following \cite{cs10} we elaborate on each of these effects below.
\begin{enumerate}
\item {\em Long-term pulse broadening function  fluctuations $\dtPBF$.}
Assuming a single ray bundle, the geometric delay due to rays at the edge of the bundle relative to the center is $\taudbar = D_{\rm eff} \theta_d^2 / 2c$.
If the scattering material has a power spectrum of fluctuations that is nearly
Kolmogorov in character \cite{cs10,ric90}, and scattering takes place in a thin screen,
then the PBF will be well approximated by $\PBF(t) = \PBFbar(t) \approx p_0 \exp{-t/\taudbar}$.
However, all of these simplifying assumptions are known to be violated at times, and so the 
PBF may vary substantially in ensemble average from this idealized form for weeks to months. 
As the large pulsar space velocity carries the LOS through the ISM, $\PBFbar(t)$ will change slowly with time, $\PBF(t) = \PBFbar(t) + \delta\PBF(t)$.
The timescale for change is dictated by the transverse size, $l_{\rm struc}$, over which 
the medium changes 
substantially in scattering properties, magnetic field structure (which affects the anisotropy level of the image \cite{rscg11,gs95,bmg+10}), or the propagation crosses filamentary or other discrete structures in the medium and is given by $T_{\rm struc} = l_{\rm struc}/V_{\rm eff}$.
The value of $\taudbar$ can be inferred in many ways, and doing this accurately is one of the main R\&D goals of interstellar mitigation groups since
this is the largest contributor to the ISM error budget after dispersion measure variations.
At the most basic level we have
\be
2\pi\taudbar \dnud = C_1,
\label{eq:uncertainty}
\ee  
where $\dnud$ is the decorrelation (diffractive) bandwidth of the spectrum \cite{cs10,cwb85}, and $C_1$ depends on the medium and its distribution along the LOS
\cite{cr98,cs10,lr00},
but is of order unity.
The long-term variation of $\dtPBF$ is not well determined along any line of sight.
However, rapid variations in DM (indicating passage through discrete structures) as well as one study of $\dtPBF$ on weeks to months timescale \cite{hs08} and studies of scattering variability in PSR~B1937+21 \cite{cwd+90,rdb+06} indicate that
 $\dtPBF \sim \taudbar$.

%In fact, substantial changes are anticipated on the refractive timescale $T_r = l_r / V_{\rm eff}$ \cite{rcb84,cr98,smc+01}, since large-scale inhomogeneiti

\item {\em Geometrical delay from angle of arrival (AOA) fluctuations  $ \dtAOA$.}
Inhomogeneities on the scale of $l_r$ cause flux variations, image distortions, and a shift in the centroid of the image \cite{cs10,rcb84}.
Although scattering in a (reference scenario) of homogeneous, isotropic scattering with
a Kolmogorv power spectrum of fluctuations produces $\theta_r \lae \theta_d$,
we know from observations that the ISM is this well behaved primarily
in the calculations of theorists and the code of modelers!
Deviations from this reference scenario are particularly relevant to long duration, 
high precision PTA observations and are summarized in \cite{rscg11} and references
therein.
% Joe Lazio improvement in the next few sentences.
If $\thetar$ is the departure of the AOA from the direct path
as viewed by an observer, the
induced time delay  is
\be
\dtAOA \approx \frac{1}{2c}
	\Deff\;
	\thetar^2
	\approx 1.21~\mu s~ \Deff(\rm kpc)\; \thetar^2({\rm mas}).
\ee
for a medium at an effective distance $D_{\mathrm{eff}}$ producing
a refraction angle of 1~mas  \cite{cs10}.
For our reference scenario, $\theta_r \lesssim \theta_d$, and $\theta_d$ can be estimated from equation (4), so that $\Delta t_{\mathrm{AOA}} \sim 160\,\mathrm{ns}\;\Delta\nu_{d,\mathrm{MHz}}^{-1}$.  
For typical PTA pulsars, $\Delta\nu_{d,\mathrm{MHz}} \sim 10\,\mathrm{MHz}$ at 1~GHz, although there is large variation around this value.
%Since $\theta_r \lae \theta_d$ for our reference scenario we can
%estimate the level of fluctuations as $\dtAOA \sim 160\, {\rm ns} / \Delta\nu_{d,\rm MHz}$, which should fluctuate on the refractive timescale $T_r$. 

\item {\em Error in correcting to the solar system barycentre (SSBC) because of angle of arrival fluctuations $\dtAOABary$.}
As discussed in \cite{lk05}, pulse arrival times need to be referred to the solar system barycentre using the exact position of the pulsar. Since AOA fluctuations offset the position, they will drive an error term that, if the position error had no time dependence of its own, would have a one year period with amplitude \cite{cs10}
\be
\dtAOABary  
	\approx 2.4~\mu s\, \thetar(\rm mas) \cos\lambda,
\ee
where $\lambda$ is the ecliptic latitude of the source. 
Fluctuations of the AOA on the refractive timescale will modulate the annual
periodicity in a stochastic fashion.

\item {\em Refractive fluctuations of the PBF $\dtPBFRISS$.}
These are conceptually distinct from the $	 \dtPBF$ variations, but
may be similar in magnitude and difficult to distinguish observationally.
As discussed in detail in \cite{cs10} these are delays connected with the
expansion and contraction of the ray bundle by refractive
scintillation. The magnitude of this effect is $\dtPBFRISS \sim \taudbar$,
but there is disagreement in the literature over whether
there is correlation or anti-correlation between $\dtPBFRISS$ and $G$,
where $G$ is the refractive gain! 
Cordes and Shannon \cite{cs10} use a ray optics argument in their 
Appendix~B and elsewhere to
find a correlation, $\tPBFRISS = G \taudbar$, 
for the simplest case of a circular symmetric image
($G = 1$ for an unrefracted image) with generalizations to non-symmetric images.
However, based on a careful wave propagation simulation,
Coles et al.\ \cite{crg+10} argue that there is an {\em anti-correlation} between
$G$ and the refractive fluctuations about the mean delay value of $\taudbar$.
There is long-standing discussion of the situation in the literature
\cite{bn85,rnb86,lrc98}, and it clearly depends (both theoretically and
observationally) on how effectively
the dominant dispersion delay is removed from the data.
Ultimately, we side with the empirical observation
of an anti-correlation between refractive flux variation and
arrival time \cite{crg+10,lrc98}.

\item {\em Diffractive fluctuations of the PBF $\dtPBFDISS$.}
The situation with $\dtPBFDISS$ is at least as complicated.
Coles et al. \cite{crg+10} state that this situation has not been
analyzed theoretically.
They argue, however, that there should also be an {\em anti-correlation}
between flux density ($G$) and arrival time. 
When the angular spectrum broadens, more power arrives
from larger angles and arrives at a later time. But, more power
is scattered out of the beam as well and, hence, the flux density
is reduced. This produces the observed anti-correlation between flux
and arrival time, operating at the diffractive scale by a different mechanism
than at the refractive scale.
There is an interesting discussion of  $\dtPBFDISS$ in \cite{cs10} (see, in particular, their
Appendix~B), but their analytical results must be treated with caution because
of the presence of effects with competing signs.
Fluctuations of the PBF on both the refractive and diffractive timescale are 
visible in figure~\ref{fig:Coles10}.
\begin{figure}[h!]
\begin{center}
      \includegraphics[scale=0.50, angle=0]{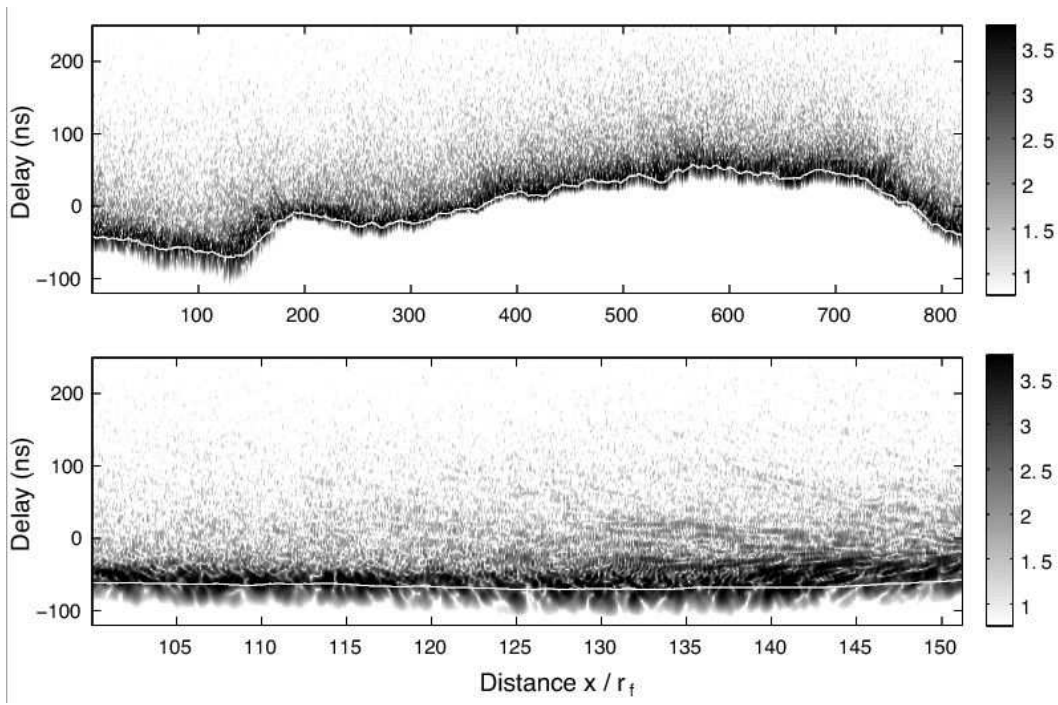}
   \caption{Pulse shape vs position for a simulation of a thin phase-changing screen
   with Kolmogorov density fluctuations\cite{crg+10}. The intensity is log10(power).
The lower panel is an expanded section of the upper one.
In units of the Fresnel scale the diffractive scale is $s_d = 0.2\; r_f$ and
the refractive scale is $s_r = 5\; r_f$ for this simulation.
The white line is the group delay through the screen at the same
transverse location as the intensity measurement.
At a transverse speed of 100~km/s the entire simulation represents a duration of
about 30~days (see the original paper for more details).	  
	\label{fig:Coles10}
   }
\end{center}
\end{figure}

\item {\em Frequency dependent variations of dispersion measure $\dtDMnu$.}
In a homogeneous medium there is only one value of the dispersion measure.
Multi-path scattering changes that, however, because
the volume over which DM is evaluated will be
frequency-dependent, growing rapidly toward lower frequencies.
In the thin screen approximation, and assuming $\theta_{\rm scatt} \propto
\nu^{-2}$, the area contributing to the DM in the screen will be
$A_{\rm scatt} \propto \nu^{-4}$.
This gives rise to another source of timing fluctuation, which 
Cordes and Shannon \cite{cs10} estimate to have an rms of 
\be
\dtDMnu
 \approx 0.12~\mu s~
	{D}^{5/6} \nu^{-23/6}
	\left( \frac{\SM}{\SMu} \right),
\ee
where $D$ is the screen-Earth distance (in kpc), $\nu$ is expressed in GHz,
and the scattering measure, which is the path integral of the magnitude
of the electron density power spectrum \cite{bcc+04}, has a value of 
${\rm SM} \approx 10^{-3.5}$~kpc~m$^{-20/3}$ for nearby pulsars, although much larger values
are possible if the LOS intercepts a region of enhanced scattering.
\end{enumerate}

%XXXXX
%\input{section_CS}
\section{Phase Reconstruction Techniques and Cyclic Spectroscopy}
The introduction of coherent dedispersion 
by Hankins \cite{hr75}
recognized the fact that dispersion is an electromagnetic phase wrapping
characterized by one number (the DM) and can be completely 
removed through the application of the exact inverse filter operating
on a pre-detection (i.e.\ proportional to the $E$-field) version of the signal.
The development of increasingly powerful signal processors has
broadened the applicability of the technique, and it has become
ubiquitous in high precision pulsar timine.
%That this technique works as well as it does is a testament to the power
%of coherent techniques in radio astronomy: the phase of the finite bandwidth e-m signal has
%been systematically rotated relative to that of a reference frequency, and 
%coherent dedispersion exactly corrects for that rotation across the band based
%on a single number, the dispersion measure.
For nearly a decade Walker and a few others have pursued the generalization
of this analysis: a coherent descattering of the signal to completely remove the
effects of multi-path scattering \cite{ws05,wksv08}.
This work had been partially successful, although the need to proceed iteratively
without direct access to the e-m phase hampered its applicability.

In 2009 Demorest introduced a new approach to the problem based
on a signal analysis technique known as {\em cyclic spectroscopy} (CS), which was 
developed by mechanical engineers studying rotating machinery, 
but not recognized as important in
radio astronomy (see references to the engineering literature in \cite{dem11}).
As the name implies
the technique is only applicable to what are known as cyclostationary signals;
fortunately, pulsar signals fall into this category. What follows is a very brief sketch of the
basics. See \cite{dem11} or the forthcoming \cite{wdv13} for more details.

%Consider complex baseband sampling of a bandwidth $B$ at the Nyquist rate
%of $\Delta = 1/B$. Assume for simplicity that there are exactly $N$ samples
%in a pulsar period. Take these $N$ time domain samples and take the signal
%into the frequency domain.
 
 Let $X(\nu)$ be the frequency domain representation of a pulsar signal.
 Then $S_x(\nu;\alpha_k)$ is the cyclic spectrum of $X(\nu)$ where
\begin{equation}
\label{eqn:fcorr}
S_x(\nu;\alpha_k) = \ev{X(\nu + \alpha_k/2) X^*(\nu - \alpha_k/2)}
\end{equation}
and $\alpha_k = k/P$, with $P$ the pulsar period, $k = 0, 1, 2, \ldots$, and the expectation value over many pulses is indicated.
If $y(t)$ is the result of passing $x(t)$
through a linear, time-invariant filter with impulse response $h(t)$
(frequency response $H(\nu)$),
\begin{equation}
y(t) = h(t) \star x(t) 
\end{equation}
\begin{equation}
Y(\nu) = H(\nu) X(\nu)
\end{equation}
then the cyclic spectra of $x$ and $y$ are related by:
\begin{equation}
\label{eqn:inout}
S_y(\nu;\alpha) = H(\nu+\alpha_k/2) H^*(\nu-\alpha_k/2) S_x(\nu;\alpha_k)
\end{equation}

If $H(\nu)$ represents the effect of the ISM on the pulsar signal, 
(\ref{eqn:inout}) shows that the phase content of $H(\nu)$ is
preserved in the cyclic spectrum, whereas in a standard power spectrum
the only information present is the filter magnitude
$\left|H(\nu)\right|^2$.  
(For simplicity we remove
the effect of dispersion from this linear filtering process and deal with it separately in
a coherent dedispersion step, so $H(\nu)$ represents the remaining ISM effects.)
This can also be understood in terms of the convolution theorem.
Pulsars, like other radio sources, emit uncorrelated broadband noise.
Modulation of the noise process by the pulsar beaming is a multiplicative
operation in the time domain. The counterpart of this is convolution in the
frequency domain. Convolution introduces correlation
into the phase structure of the signal.
Fourier complementarity means that a pulse width of $W$ results in a correlation of
the phase over a frequency interval $\sim 1/W$.
This correlated phase (within one pulse) is retained when it passes through
$H(\nu)$, and the cyclic spectrum makes the phase of $H(\nu)$
recoverable through
the conjugate, shift, and multiply nature of (\ref{eqn:inout}).
Exactly how to recover $\Phi_H$ from  (\ref{eqn:inout}) is a
substantial subject in itself and is the focus of the forthcoming  paper \cite{wdv13}.
Once the amplitude and phase of $H(\nu)$ have been recovered, a 
coherent descattering of the signal is straightforward in principle.
%This is a fundamental difference between the
%two approaches that will enable most of the novel applications described
%in \S\ref{sec:apps}.

\begin{figure}[htb]
\begin{center}
      \includegraphics[scale=0.90, angle=0]{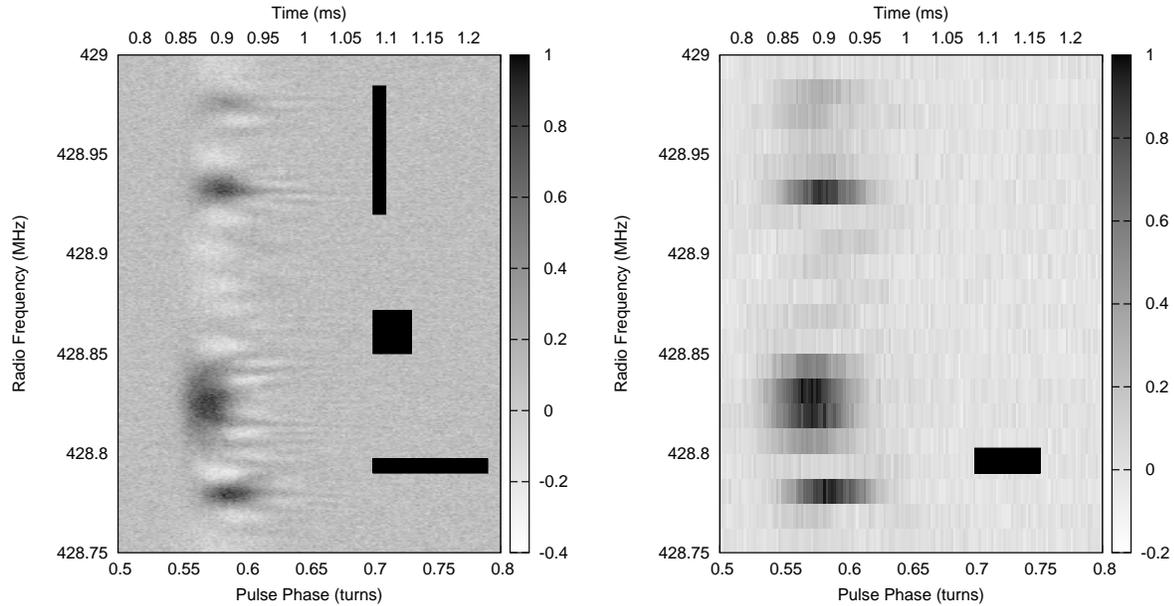}
   \caption{From Demorest 2011 \cite{dem11}.
        Left: Detail of periodic spectrum of B1937+21.  
        The black rectangles each contain unity time-frequency area.
        Unlike a traditional dynamic spectrum the periodic spectrum
        incorporates e-m phase information resulting in positive(dark)
        and negative (light) regions on fine scales. See the original
        article for details.
        Right: The same data processed with a standard coherently
        dedispersed filter bank using a frequency resolution of 
        12.5~kHz and corresponding intrinsic time resolution of 80~$\mu$s (black rectangle).
	See \cite{dem11} for more details.
	\label{fig:demCS1937}
   }
\end{center}
\end{figure}

%theory\\
%application\\
The above basics represent the more ambitious use of CS in pulsar
signal processing, but CS has an important advantage over standard
techniques even if coherent descattering is not the goal.
This is illustrated graphically in figure~\ref{fig:demCS1937} \cite{dem11}.
This shows a dynamic spectrum produced
with CS (left) and traditional techniques (right).
By coherently using a much longer data span than is possible with standard
techniques, CS can achieve a much finer time and
frequency resolution, which can be especially valuable when observing
millisecond pulsars with substantial scattering, hence  values
of $\Delta\nu_d \ll 1$~MHz.
Recently Jones in the NANOGrav collaboration has developed a
real-time CS capability based on similar reconfigurable digital
hardware as PUPPI and GUPPI, which is available at Arecibo and Green
Bank \cite{jdv13}.
This is being put into routine operation for supplemental data-taking
in parallel for each of our standard timing runs.

%XXXXX
%\input{section_mitigation}
\section{Mitigation Techniques}
\label{sec:mitigation}
The comprehensive arrival time model presented in \cite{cs10} can be expressed
as
\be
t_{\nu} = t_{\infty} + t_{\DM}(\nu) + t_C(\nu) + \tWHITE(\nu), 
\label{eq:tsimple}
\ee 
where $t_{\infty}$ is the arrival time at infinite frequency, and the other terms
represent, respectively, DM fluctuations, other chromatic (primarily scattering related) terms,
and a white noise contribution made up of radiometer noise and pulse jitter noise
due to the finite number of pulses included in any observation.
Cordes and Shannon consider two broad approaches to the mitigation of ISM effects:
i) correct for DM variations, but ignore refractive and multi-path
scattering effects and ii) use multi-wavelength (and broadband) 
information to aggressively remove all chromatic effects.
They lay out a detailed plan of attack in both approaches, but
the test of strategy against empirical performance is in its
infancy.
We refer the reader to \cite{cs10} as an excellent
starting point in developing a comprehensive mitigation
strategy and encourage the PTAs to intercompare results
in this important arena.

%XXXXX
%\input{section_future}
\section{Future Developments}
\label{sec:future}
Very few of the areas outlined above are settled at the
level of precision needed for successful PTA timing efforts,
particularly when we consider the multi-year to multi-decade duration of those
observations.
R\&D will continue in all of these areas with various PTAs taking
the lead on issues of particular importance to them and
the IPTA providing structure to sort out optimal solutions.

PTA-based science produces
numerous ancillary projects of great interest to communities outside
of gravitational studies.
The data we are taking, in both its precision and scope, will be of
enormous benefit to the study of the ionized interstellar medium
on a large range of size scales.
Subgroups within the three PTA collaborations pursue this
ancillary science, and we can expect many exciting results over
the next decade.

Looking ahead on a decade or longer timescale, several 
researchers have highlighted the enormous benefits (particularly in terms of {\em greatly} improved angular resolution in the localization of a GW source)
to accurately determining the length
of one or more ``arms" of the Galactic scale interferometer
the IPTA is constructing \cite{bp12}.
The requisite accuracy is a fraction of the gravitational wavelength
involved, so we might set $\pm 1$~ly as a reference goal.
Recent high-precision VLBI work \cite{dvtb08,dbl+13} is approaching
this goal for some pulsars, and this is likely to serve as an important
element in bootstrapping to the needed precision.
Building on ideas and results of the last several decades 
\cite{crsc06,smc+01,shm+03,hsa+05,ps06a,wmsz04},     % earlyscintarcstudies
Pen (CITA) has
been championing the use of interstellar scattering as a powerful
tool to determine pulsar distances to the needed accuracy, 
particularly by including the use of VLBI 
techniques \cite{pmdb13}.

\begin{figure}[h!]
  \begin{center}
     \hfill \includegraphics[scale=0.35, angle=0]{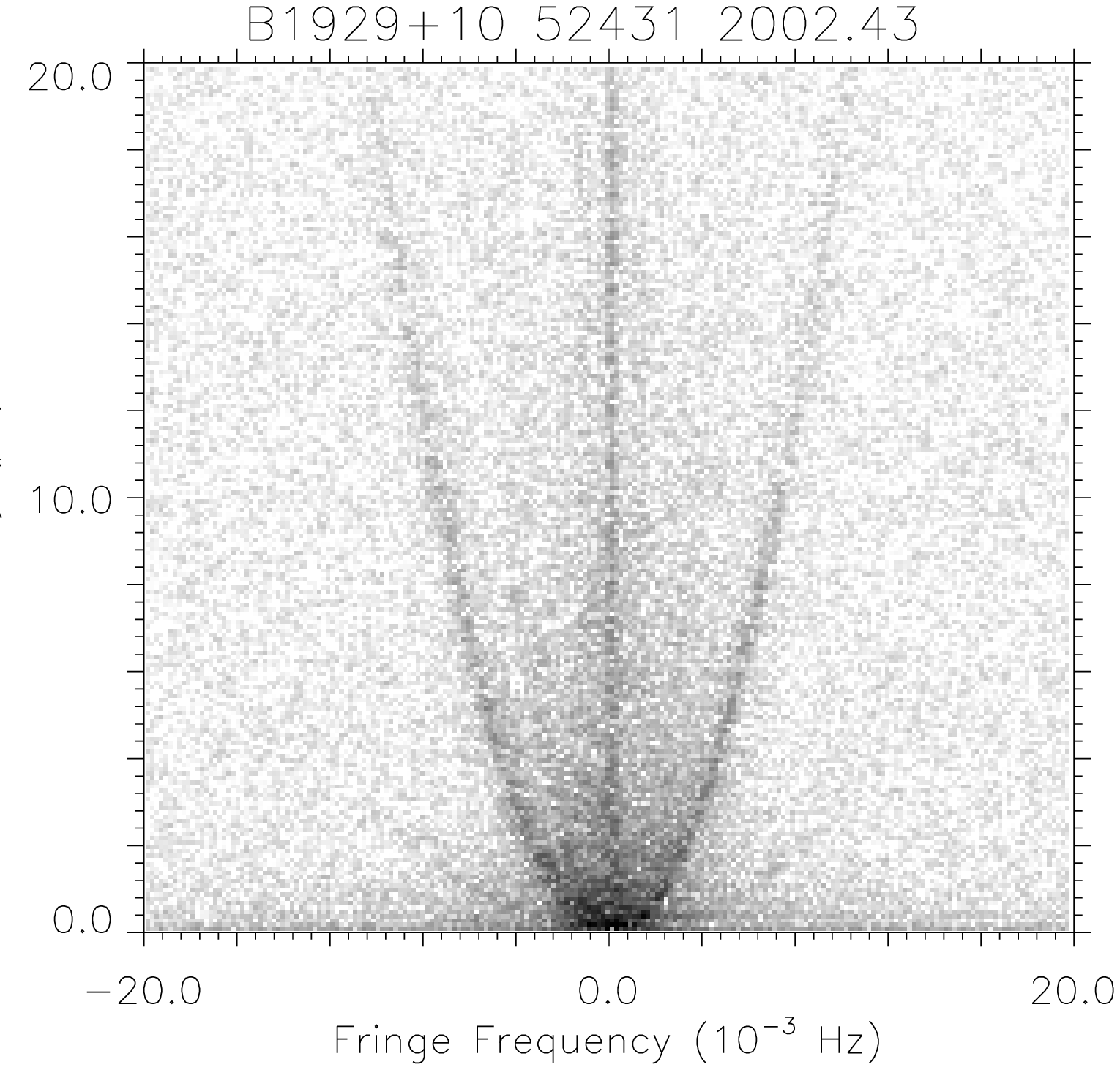} % 
      \hfill \includegraphics[scale=0.45, angle=0]{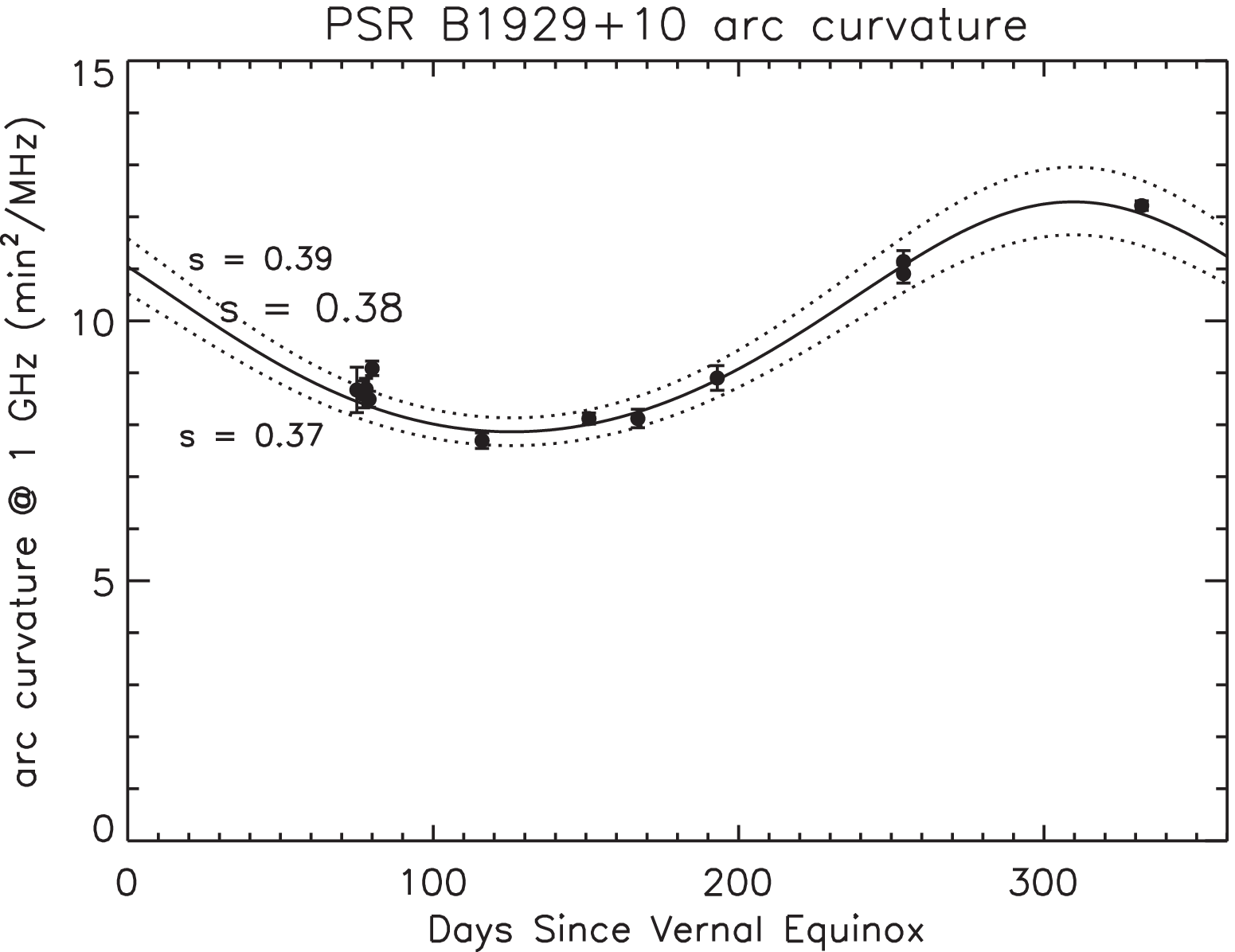} \hspace*{\fill}
   \caption{(Left) A secondary spectrum for B1929+10 obtained in a one hour observation
   with the Arecibo telescope \cite{crsc06}. The sharply delineated parabolic (scintillation)
   arc is due to scattering in a highly localized screen along the LOS and also requires
   an anisotropically scattered image \cite{crsc06,rscg11}. Inspection of this and secondary spectra from other epochs shows the presence of two other scintillation arcs, which vary in prominence over multiple years \cite{ps06a}.
   (Right) An analysis of scintillation arcs like the one on the left, along with estimates of the pulsar distance and its measured proper motion, allow the scattering screen to be located, expressed here as a fraction of the distance from the pulsar to the Earth. 
 See \cite{crsc06} for details and caveats. This figure originally appeared in \cite{shr05}.
  	\label{fig:1929earthmotion}
   }
  \end{center}
\end{figure}

Using a single-dish example we underscore the potential of high precision astrometry employing multi-path scattering from AU-sized structures in the ISM.
In figure~\ref{fig:1929earthmotion} we show that the scintillation
arc(s) for a nearby, lightly scattered pulsar can be remarkably sharp.
Although this has not been fully explored yet for the millisecond
pulsars in typical PTAs, there is no doubt that some pulsars in
the sample -- perhaps even particularly important ones from a
timing precision/sensitiviy point of view -- will have equally sharp arcs
because of scattering over a very small range of the LOS to
the pulsar.
The curvature of a scintillation arc is a sensitive function of
the relative velocities involved \cite{crsc06}. When the Earth's motion is
large compared to the pulsar's motion, it is possible to see the
effect of that motion as a function of time of year. This is shown
in the right hand panel of figure~\ref{fig:1929earthmotion} \cite{shr05}.

Overlaid on the data points are three curves for the location of the
scattering screen as a function of the fractional distance from the
pulsar to the Earth.
In this case we were able to determine the fractional distance
to $\pm 1$~\%.
Although this is neither the required quantity (distance to the pulsar),
nor the requisite precision, nor an absolute (versus a relative) measurement,
this and other work \cite{bmg+10} hints at the enormous potential
for ultra-high precision astrometry using pulsar scintillation with large
telescopes, particularly when linked using VLBI techniques.

%XXXXX
\section{Summary}
\label{sec:conc}
Based on the discussion above and similar research
we emphasize the following points:
\begin{enumerate}
\item Individual lines of sight to important PTA pulsars must be monitored and mitigation strategies tailored to the ISM conditions along that LOS. Multi-wavelength and broadband techniques are crucial to this effort.
\item Accurate and careful correction for dispersion measure variations is an essential
component to successful PTA timing efforts and is not yet a settled topic despite
important recent contributions.
\item Multi-path scattering introduces a myriad of effects at the several microsecond level
and below. These need to be determined and mitigated through the use of
multi-frequency or ultra-broadband observations and effective algorithms.
\item Cyclic spectroscopy and other phase retrieval techniques are likely to
play an increasingly important role in interstellar mitigation efforts.
\item Pulsar scintillation may provide a crucial tool in determining high precision
distances to pulsars, greatly enhancing the characteristics of PTAs as gravitational
wave detectors.
\end{enumerate}

\ack
We thank the members of the NANOGrav interstellar medium mitigation (IMM) working  group for their comments and contributions as well as important suggestions from NANOGrav members Justin Ellis, Joseph Lazio, and David Nice. We gratefully acknowledge the support of the NSF through AAG grant 1009580 and PIRE award number 0968126. Comments from both referees were valuable in improving the manuscript.

%\section*{References}
%\bibliographystyle{iopart-num} 
%\bibliography{journals_apj,scintillation,modrefs,psrrefs,crossrefs}

\newpage
\section*{References}
\begin{thebibliography}{10}
\expandafter\ifx\csname url\endcsname\relax
  \def\url#1{{\tt #1}}\fi
\expandafter\ifx\csname urlprefix\endcsname\relax\def\urlprefix{URL }\fi
\providecommand{\eprint}[2][]{\url{#2}}

\bibitem{lk05}
Lorimer D~R and Kramer M 2005 {\em Handbook of Pulsar Astronomy\/} (Cambridge
  University Press)

\bibitem{dfg+13}
{Demorest} P~B, {Ferdman} R~D, {Gonzalez} M~E, {Nice} D, {Ransom} S, {Stairs}
  I~H, {Arzoumanian} Z, {Brazier} A, {Burke-Spolaor} S, {Chamberlin} S~J,
  {Cordes} J~M, {Ellis} J, {Finn} L~S, {Freire} P, {Giampanis} S, {Jenet} F,
  {Kaspi} V~M, {Lazio} J, {Lommen} A~N, {McLaughlin} M, {Palliyaguru} N,
  {Perrodin} D, {Shannon} R~M, {Siemens} X, {Stinebring} D, {Swiggum} J and
  {Zhu} W~W 2013 {\em ApJ\/} {\bf 762} 94

\bibitem{kcs+13}
Keith M~J, Coles W, Shannon R~S and et~al 2012 {\em MNRAS\/} {\bf 429} 2161

\bibitem{hs08}
Hemberger D~A and Stinebring D~R 2008 {\em ApJ\/} {\bf 674} L37--L40

\bibitem{cr98}
{Cordes} J~M and {Rickett} B~J 1998 {\em ApJ\/} {\bf 507} 846--860

\bibitem{crsc06}
{Cordes} J~M, {Rickett} B~J, {Stinebring} D~R and {Coles} W~A 2006 {\em ApJ\/}
  {\bf 637} 346--365

\bibitem{smc+01}
{Stinebring} D~R, {McLaughlin} M~A, {Cordes} J~M, {Becker} K~M, {Goodman}
  J~E~E, {Kramer} M~A, {Sheckard} J~L and {Smith} C~T 2001 {\em ApJ\/} {\bf
  549} L97--L100

\bibitem{shm+03}
{Stinebring} D~R, {Hill} A~S, {McLaughlin} M~A, {Becker} K~M, {Cordes} J~M and
  {Kramer} M 2003 {\em Radio Pulsars\/} ({\em Astronomical Society of the
  Pacific Conference Series\/} vol 302) ed {Bailes} M, {Nice} D~J and
  {Thorsett} S~E p 263

\bibitem{hsb+03}
Hill A~S, Stinebring D~R, Barnor H~A, Berwick D~E and Webber A~B 2003 {\em
  ApJ\/} {\bf 599} 457--464

\bibitem{hsa+05}
{Hill} A~S, {Stinebring} D~R, {Asplund} C~T, {Berwick} D~E, {Everett} W~B and
  {Hinkel} N~R 2005 {\em ApJL\/} {\bf 619} L171--L174

\bibitem{pmsr05}
{Putney} M~L, {Minter} A~H, {Stinebring} D~R and {Ransom} S~M 2005 {\em
  American Astronomical Society Meeting Abstracts\/} ({\em Bulletin of the
  American Astronomical Society\/} vol~37) p 183.15

\bibitem{ps06a}
{Putney} M~L and {Stinebring} D~R 2006 {\em Chin. J. Atron. Astrophys., Suppl.
  2\/} {\bf 6} 233--236

\bibitem{sti07}
{Stinebring} D 2007 {\em SINS - Small Ionized and Neutral Structures in the
  Diffuse Interstellar Medium\/} ({\em Astronomical Society of the Pacific
  Conference Series\/} vol 365) ed {Haverkorn} M and {Goss} W~M p 254

\bibitem{hs07}
{Heiles} C and {Stinebring} D 2007 {\em SINS - Small Ionized and Neutral
  Structures in the Diffuse Interstellar Medium\/} ({\em Astronomical Society
  of the Pacific Conference Series\/} vol 365) ed {Haverkorn} M and {Goss} W~M
  p 331

\bibitem{rscg11}
{Rickett} B, {Stinebring} D, {Coles} B and {Gao} J 2011 {\em American Institute
  of Physics Conference Series\/} ({\em American Institute of Physics
  Conference Series\/} vol 1357) ed {Burgay} M, {D'Amico} N, {Esposito} P,
  {Pellizzoni} A and {Possenti} A pp 97--100

\bibitem{cs10}
{Cordes} J~M and {Shannon} R~M 2010 {\em arXiv:1010.3785\/}

\bibitem{ric90}
Rickett B~J 1990 {\em Ann. Rev. Astr. Ap.\/} {\bf 28} 561--605

\bibitem{rcb84}
Rickett B~J, Coles W~A and Bourgois G 1984 {\em A\&A\/} {\bf 134} 390--395

\bibitem{ssh+00}
Stinebring D~R, Smirnova T~V, Hankins T~H, Hovis J, Kaspi V, Kempner J, Meyers
  E and Nice D~J 2000 {\em ApJ\/} {\bf 539} 300--316

\bibitem{ebf+70}
{Ewing} M~S, {Batchelor} R~A, {Friefeld} R~D, {Price} R~M and {Staelin} D~H
  1970 {\em ApJL\/} {\bf 162} L169

\bibitem{ra82}
Roberts J~A and Ables J~G 1982 {\em MNRAS\/} {\bf 201} 1119--1138

\bibitem{hwg85}
Hewish A, Wolszczan A and Graham D~A 1985 {\em MNRAS\/} {\bf 213} 167

\bibitem{cw86}
Cordes J~M and Wolszczan A 1986 {\em ApJ\/} {\bf 307} L27--L32

\bibitem{wc87}
Wolszczan A and Cordes J~M 1987 {\em ApJ\/} {\bf 320} L35--L39

\bibitem{rlg97}
Rickett B~J, Lyne A~G and Gupta Y 1997 {\em MNRAS\/} {\bf 287} 739--752

\bibitem{lr99}
{Lambert} H~C and {Rickett} B~J 1999 {\em ApJ\/} {\bf 517} 299--317

\bibitem{nar92a}
Narayan R 1992 {\em Phil. Trans. Roy. Soc. A\/} {\bf 341} 151--165

\bibitem{lgs12}
Lyne A and Graham-Smith F 2012 {\em Pulsar Astronomy\/} (Cambridge University
  Press)

\bibitem{es04}
{Elmegreen} B~G and {Scalo} J 2004 {\em ARAA\/} {\bf 42} 211--273

\bibitem{se04}
{Scalo} J and {Elmegreen} B~G 2004 {\em ARAA\/} {\bf 42} 275--316

\bibitem{yhc+07}
{You} X~P, {Hobbs} G, {Coles} W~A, {Manchester} R~N, {Edwards} R, {Bailes} M,
  {Sarkissian} J, {Verbiest} J~P~W, {van Straten} W, {Hotan} A, {Ord} S,
  {Jenet} F, {Bhat} N~D~R and {Teoh} A 2007 {\em MNRAS\/} {\bf 378} 493

\bibitem{vl13}
van Haasteren R and Levin Y 2013 {\em MNRAS\/} {\bf 428} 1147

\bibitem{esv13}
Ellis J~A, Siemens X and van Haasteren R 2013 {\em ApJ\/} {\bf 769} 63

\bibitem{fb90}
Foster R~S and Backer D~C 1990 {\em ApJ\/} {\bf 361} 300

\bibitem{cfrc87}
Coles W~A, Frehlich R~G, Rickett B~J and Codona J~L 1987 {\em ApJ\/} {\bf 315}
  666--674

\bibitem{crg+10}
{Coles} W~A, {Rickett} B~J, {Gao} J~J, {Hobbs} G and {Verbiest} J~P~W 2010 {\em
  ApJ\/} {\bf 717} 1206--1221

\bibitem{gs95}
Goldreich P and Sridhar S 1995 {\em ApJ\/} {\bf 438} 763--775

\bibitem{bmg+10}
{Brisken} W~F, {Macquart} J~P, {Gao} J~J, {Rickett} B~J, {Coles} W~A, {Deller}
  A~T, {Tingay} S~J and {West} C~J 2010 {\em ApJ\/} {\bf 708} 232--243

\bibitem{cwb85}
Cordes J~M, Weisberg J~M and Boriakoff V 1985 {\em ApJ\/} {\bf 288} 221--247

\bibitem{lr00}
{Lambert} H~C and {Rickett} B~J 2000 {\em ApJ\/} {\bf 531} 883--901

\bibitem{cwd+90}
Cordes J~M, Wolszczan A, Dewey R~J, Blaskiewicz M and Stinebring D~R 1990 {\em
  ApJ\/} {\bf 349} 245

\bibitem{rdb+06}
{Ramachandran} R, {Demorest} P, {Backer} D~C, {Cognard} I and {Lommen} A 2006
  {\em ApJ\/} {\bf 645} 303--313

\bibitem{bn85}
Blandford R~D and Narayan R 1985 {\em MNRAS\/} {\bf 213} 591--611

\bibitem{rnb86}
Romani R~W, Narayan R and Blandford R 1986 {\em MNRAS\/} {\bf 220} 19--49

\bibitem{lrc98}
Lestrade J, Rickett B~J and I C 1998 {\em A\&A\/} {\bf 334} 1068

\bibitem{bcc+04}
{Bhat} N~D~R, {Cordes} J~M, {Camilo} F, {Nice} D~J and {Lorimer} D~R 2004 {\em
  ApJ\/} {\bf 605} 759--783

\bibitem{hr75}
{Hankins} T~H and {Rickett} B~J 1975 {\em Methods in Computational Physics
  Volume 14 --- Radio Astronomy\/} (New York: Academic Press) pp 55--129

\bibitem{ws05}
{Walker} M~A and {Stinebring} D~R 2005 {\em MNRAS\/} {\bf 362} 1279--1285

\bibitem{wksv08}
{Walker} M~A, {Koopmans} L~V~E, {Stinebring} D~R and {van Straten} W 2008 {\em
  MNRAS\/} {\bf 388} 1214--1222

\bibitem{dem11}
{Demorest} P~B 2011 {\em MNRAS\/} {\bf 416} 2821--2826

\bibitem{wdv13}
Walker M~A, Demorest P~B and van Straten W 2013 {\em ApJ {\rm submitted}\/}

\bibitem{jdv13}
Jones G, Demorest P~B and van Straten W 2013 {\em \rm in preparation\/}

\bibitem{bp12}
{Boyle} L and {Pen} U~L 2012 {\em PhysRevD\/} {\bf 86} 124028

\bibitem{dvtb08}
{Deller} A~T, {Verbiest} J~P~W, {Tingay} S~J and {Bailes} M 2008 {\em ApJ\/}
  {\bf 685} L67--L70

\bibitem{dbl+13}
{Deller} A~T, {Boyles} J, {Lorimer} D~R, {Kaspi} V~M, {McLaughlin} M~A,
  {Ransom} S, {Stairs} I~H and {Stovall} K 2013 {\em ApJ\/} {\bf 770} 145

\bibitem{wmsz04}
{Walker} M~A, {Melrose} D~B, {Stinebring} D~R and {Zhang} C~M 2004 {\em
  MNRAS\/} {\bf 354} 43--54

\bibitem{pmdb13}
{Pen} U~L, {Macquart} J~P, {Deller} A and {Brisken} W 2013 {\em ArXiv
  1301.7505\/}

\bibitem{shr05}
{Stinebring} D~R, {Hill} A~S and {Ransom} S~M 2005 {\em Binary Radio Pulsars\/}
  ({\em Astronomical Society of the Pacific Conference Series\/} vol 328) ed
  {Rasio} F~A and {Stairs} I~H p 349

\end{thebibliography}

\end{document}